\documentclass[a4paper,aps,preprint,groupedaddress, showkeys]{revtex4}
\usepackage{amssymb}
\usepackage{amsmath}
\usepackage{graphicx}
\usepackage[T2A]{fontenc}

\begin{document}

\title{Gravitational frequency shift of light signals in a pulsating dark matter halo}
\author{Vladimir A. Koutvitsky}
\email{vak@izmiran.ru}
\author{Eugene M. Maslov}
\email{zheka@izmiran.ru}
\affiliation{Pushkov Institute of Terrestrial Magnetism, Ionosphere and Radio Wave Propagation RAS, Moscow, Troitsk, Russia}

\begin{abstract}

The gravitational frequency shift of light signals from the center of a
spherically symmetric non-static matter distribution is considered. 
Explicit formulas for the ratio of the emitted and received frequencies are
obtained in the case of the oscillating scalar dark matter with logarithmic potential.

\end{abstract}
\keywords{dark matter halo, oscillating scalar field, non-static metrics, red shift, blue shift.}

\maketitle

Modern observations show that approximately one fourth of the total mass of
the Universe falls on dark matter, which forms galactic halos and, possibly,
separate self-gravitating lumps \cite{Fr}. These results are based on the
study of motion of visible matter and propagation of photons in a gravitational field
created by various spatial distributions of dark matter. If the distribution
of dark matter is non--static, then, generally speaking, the corresponding
metric of space--time will be non-static too. In this paper, we investigate
the gravitational frequency shift of a light signal propagating in a
spherically symmetric pulsating lump of scalar dark matter. The
corresponding metric will be

\begin{equation}
ds^{2}=B(t,r)\,dt^{2}-A(t,r)\,dr^{2}-r^{2}(d\vartheta ^{2}+\sin
^{2}\vartheta \,d\varphi ^{2}).  \label{eq1}
\end{equation}%
The general formula for the frequency shift of a photon propagating in a
gravitational field is well known (see, e.g., \cite{Schrodinger}). Namely,
the ratio of the source's proper frequency at the photon emission point to
the photon frequency at the observation point is equal to the ratio of the
scalar product of the photon's 4--momentum at the emission point and of the
source 4-velocity to the scalar product of the photon's 4--momentum at the
observation point and of the observer 4--velocity. The 4--momentum of the
photon satisfies the null--geodesic equation, which in the case of a
non-static metric reduces to a rather complicated system of non--linear
differential equations with respect to the affine parameter. Analytical
solutions of this system can be obtained only in certain particular cases,
for example, for the metrics describing some expanding isotropic spacetimes \cite%
{Schrodinger}, plane gravitational waves \cite{Kauf, Far}, and small
fluctuations of the gravitational potential \cite{Ford, Thom}. In the case
of the spherically symmetric metric (\ref{eq1}) ,
the geodesic equation for the photon 4--momentum was solved numerically in the paper \cite{Bosk}.

Note that it is possible to obtain a simple formula for the frequency shift
in that case when the source is located at the center of a spherically
symmetric distribution, and the observer is at rest at a point $r=R$.
Indeed, setting in (\ref{eq1}) $ds=d\vartheta =d\varphi =0$, we find the
equation for the photon trajectory in the form%
\begin{equation}
dt/dr=\sqrt{A(t,r)/B(t,r)}.  \label{eq2}
\end{equation}%
We now consider two light pulses that are successively emitted at the point $%
r=0$ through a small time interval $\delta t(0)$. The corresponding close
trajectories, $t(r)$ and $t(r)+\delta t(r)$, satisfy the equation (\ref{eq2}%
) on the interval $0\leqslant r\leqslant R$. Substituting $t(r)+\delta t(r)$
into (\ref{eq2}) and expanding the right--hand side near the trajectory $t(r)$%
, we find the ratio $\delta t(r)/\delta t(0)$. As a result, for the ratio of
the proper frequency of the source $\nu _{0}$ to the observed frequency $\nu
_{R}$, we obtain%
\begin{equation}
\frac{\nu _{0}}{\nu _{R}}=\sqrt{\frac{B(t(R),R)}{B(t(0),0)}}\exp \int_{0}^{R}%
\left[ \frac{\partial }{\partial t}\sqrt{\frac{A(t,r)}{B(t,r)}}\right]
_{t=t(r)}dr.  \label{eq3}
\end{equation}%
Note that this formula is also valid in the comoving coordinates: in this case,
in (\ref{eq3}), we must put $B=1$. For example, for the Friedmann --
Robertson -- Walker metric from (\ref{eq3}) it immediately follows that the
ratio $\nu _{0}/\nu _{R}$ is equal to the ratio of the scale factors taken
at the moments $t(R)$ and $t(0)$, respectively.

It is easy to verify that the formula (\ref{eq3}) can be represented in the
following equivalent form%
\begin{equation}
\frac{\nu _{0}}{\nu _{R}}=\sqrt{\frac{A(t(R),R)}{A(t(0),0)}}\exp
\int_{t(0)}^{t(R)}\left[ \frac{\partial }{\partial r}\sqrt{\frac{B(t,r)}{%
A(t,r)}}\right] _{r=r(t)}dt.  \label{eq4}
\end{equation}%
In the post--Newtonian approximation, the formula obtained in \cite{Khmel}
follows from this.

Using these results, we calculate the gravitational frequency shift of the
light signal from the center of the self--gravitating pulsating lump of a
real scalar field $\phi (t,r)$ with potential%
\begin{equation}
U(\phi )=\frac{1}{2}m^{2}\phi ^{2}\left( 1-\ln \frac{\phi ^{2}}{\sigma ^{2}}%
\right) ,  \label{eq5}
\end{equation}%
where $\sigma $ is the characteristic magnitude of the scalar field, $m$ is
the mass (in units $\hbar =c=1$). Potentials of this kind arise in the
inflationary cosmology \cite{Bar}, as well as in some supersymmetric
extensions of the standard model \cite{Enq}. The corresponding solution of
the Einstein--Klein--Gordon system, describing the self-gravitating pulson,
was found in \cite{Kout} in the weak field approximation. It has the form:%
\begin{equation}
\phi (t,r)/\sigma =[a(\theta )+\varkappa Q(\theta ,\rho )+O(\varkappa
^{2})]e^{(3-\rho ^{2})/2},  \label{eq6}
\end{equation}%
\begin{equation}
A(t,r)=(1-\rho _{g}/\rho )^{-1},\quad B(t,r)=(1-\rho _{g}/\rho )e^{-s},
\label{eq7}
\end{equation}%
\begin{equation}
\rho _{g}(\tau ,\rho )=\varkappa \lbrack V_{\max }((\sqrt{\pi }/2)e^{\rho
^{2}}\mathrm{erf}\rho -\rho )-a^{2}\rho ^{3}]e^{3-\rho ^{2}}+O(\varkappa
^{2}),  \label{eq8}
\end{equation}%
\begin{equation}
s(\tau ,\rho )=\varkappa (2V_{\max }+a^{2}\ln a^{2}+a^{2}\rho ^{2})e^{3-\rho
^{2}}+O(\varkappa ^{2}),  \label{eq9}
\end{equation}%
where $a(\theta (\tau ))$ satisfies the equation of a non--linear oscillator,%
\begin{equation}
a_{\theta \theta }=-dV/da,\quad V(a)=(a^{2}/2)(1-\ln a^{2}),  \label{eq10}
\end{equation}%
$\tau =mt$, $\rho =mr$, $\varkappa =4\pi G\sigma ^{2}\ll 1$ ($G$ is the
gravitational constant), $V_{\max }=V(a_{\max })$, $\theta _{\tau
}=1+\varkappa \Omega +O(\varkappa ^{2})$, $\varkappa \Omega $ is the pulson
frequency correction. The first term in the formula (\ref{eq6}) describes
the anharmonic oscillations $-a_{\max }\leqslant a(\theta )\leqslant a_{\max
}$ in the symmetric potential $V(a)$. The function $Q(\theta ,\rho )$ is a
series of Hermite polynomials whose coefficients are periodic (in $\theta $)
solutions of the nonhomogeneous Hill equations. In the paper \cite{Kout},
initial conditions and the correction $\varkappa \Omega $ were found for
which such solutions exist. The stability of these solutions essentially
depends on the amplitude of oscillations $a_{\max }$. It turned out, that in
some intervals of $a_{\max }$ values, solutions with high accuracy retain
their periodicity, making hundreds of oscillations. Similar quasi--stability
intervals were also found when studying the effect of external perturbations
on a pulson in the absence of gravity \cite{Kout2, Kout3}. Since to
calculate the gravitational frequency shift we need to know only the metric
coefficients $A(t,r)$ and $B(t,r)$, we will not discuss in more detail here
the structure of the function $Q(\theta ,\rho )$, whose influence on the
metric manifests itself in the next orders in $\varkappa $, but just assume
that $a_{\max }$ belongs to one of the quasi--stability intervals.

Let the function $a(\theta )$ be an even periodic solution of the equation (%
\ref{eq10}) with a period%
\begin{equation}
T(a_{\max })=2\pi /\omega =4\int_{0}^{1}\frac{d\xi }{\sqrt{(1-\ln a_{\max
}^{2})(1-\xi ^{2})+\xi ^{2}\ln \xi ^{2}}}.  \label{eq11}
\end{equation}%
Then%
\begin{eqnarray}
a^{2} &=&A_{0}/2+\sum_{n=1}^{\infty }A_{n}\cos 2n\omega \theta ,
\label{eq12} \\
a^{2}\ln a^{2} &=&C_{0}/2+\sum_{n=1}^{\infty }C_{n}\cos 2n\omega \theta .
\label{eq13}
\end{eqnarray}%
Using the equations (\ref{eq10}), it is easy to show that the Fourier
coefficients $A_{n}$ and $C_{n}$ are related by%
\begin{equation}
C_{0}=\frac{A_{0}}{2}-2V_{\max },\quad C_{n}=\frac{A_{n}}{2}[1-2(n\omega
)^{2}],  \label{eq14}
\end{equation}%
where%
\begin{equation}
A_{0}=\frac{8a_{\max }^{2}}{T(a_{\max })}\int_{0}^{1}\frac{\xi ^{2}d\xi }{%
\sqrt{(1-\ln a_{\max }^{2})(1-\xi ^{2})+\xi ^{2}\ln \xi ^{2}}},  \label{eq15}
\end{equation}%
\begin{equation}
A_{n}=\frac{8}{T}\int_{0}^{T/4}a^{2}(\theta )\cos 2n\omega \theta \;d\theta .
\label{eq16}
\end{equation}

We now return to the equations (\ref{eq2}), (\ref{eq3}). Note that due to
the smallness of $\varkappa $, the right-hand side of the equation (\ref{eq2}%
) is close to unity, so that its time derivative in the equation (\ref{eq3})
is of the order $O(\varkappa )$. Therefore, when calculating the integral in
(\ref{eq3}), we can take the photon trajectory in the form $t(r)=r+t(0)$.
Thus, substituting (\ref{eq7})-(\ref{eq9}) into the formula (\ref{eq3}) and
using (\ref{eq12})-(\ref{eq14}), we can set $\theta \approx \tau (\rho
)=\rho +\tau (0)$, where $\tau (0)=\tau (\mathcal{R})-\mathcal{R}$, $%
\mathcal{R}=mR$. As a result, we finally get%
\[
\frac{\nu _{0}}{\nu _{R}}=1+z\approx 1+\varkappa \frac{e^{3}}{2}\bigg{\{}%
V_{\max }\left( 1-\frac{\sqrt{\pi }}{2}\frac{\mathrm{erf}\mathcal{R}}{%
\mathcal{R}}\right) +\frac{A_{0}}{4}\left( 1-e^{-\mathcal{R}^{2}}\right) 
\]%
\begin{equation}
+\sum_{n=1}^{\infty }\left[ C_{n}\left( \cos 2n\omega \tau (0)-e^{-\mathcal{R%
}^{2}}\cos 2n\omega \tau (\mathcal{R})\right) -2n\omega P_{n}(\tau (0),%
\mathcal{R})\right] \bigg{\}},  \label{eq17}
\end{equation}%
where%
\begin{eqnarray}
P_{n}(\tau (0),\mathcal{R}) &=&\big{[}\,C_{n}I_{1}(n\omega ,\mathcal{R}%
)-A_{n}I_{3}(n\omega ,\mathcal{R})\,\big{]}\cos 2n\omega \tau (0)  \nonumber
\\
&&+\big{[}\,C_{n}I_{2}(n\omega ,\mathcal{R})-A_{n}I_{4}(n\omega ,\mathcal{R}%
)\,\big{]}\sin 2n\omega \tau (0),  \label{eq18}
\end{eqnarray}%
\begin{eqnarray}
I_{1}(n\omega ,\mathcal{R}) &=&\int_{0}^{\mathcal{R}}e^{-\rho ^{2}}\sin
2n\omega \rho \;d\rho ,  \nonumber \\
I_{2}(n\omega ,\mathcal{R}) &=&\int_{0}^{\mathcal{R}}e^{-\rho ^{2}}\cos
2n\omega \rho \;d\rho ,  \nonumber \\
I_{3}(n\omega ,\mathcal{R}) &=&\int_{0}^{\mathcal{R}}\rho ^{2}e^{-\rho
^{2}}\sin 2n\omega \rho \;d\rho ,  \nonumber \\
I_{4}(n\omega ,\mathcal{R}) &=&\int_{0}^{\mathcal{R}}\rho ^{2}e^{-\rho
^{2}}\cos 2n\omega \rho \;d\rho .  \label{eq19}
\end{eqnarray}%
Integrating by parts, it is easy to show that%
\begin{eqnarray}
I_{3}(n\omega ,\mathcal{R}) &=&\frac{n\omega }{2}+\frac{1}{2}\big{[}%
1-2(n\omega )^{2}\big{]}I_{1}(n\omega ,\mathcal{R})  \nonumber \\
&&-\frac{1}{2}e^{-\mathcal{R}^{2}}\left( n\omega \cos 2n\omega \mathcal{R}+%
\mathcal{R}\sin 2n\omega \mathcal{R}\right) ,  \nonumber \\
I_{4}(n\omega ,\mathcal{R}) &=&\frac{1}{2}\big{[}1-2(n\omega )^{2}\big{]}%
I_{2}(n\omega ,\mathcal{R})  \nonumber \\
&&+\frac{1}{2}e^{-\mathcal{R}^{2}}\left( n\omega \sin 2n\omega \mathcal{R}-%
\mathcal{R}\cos 2n\omega \mathcal{R}\right) .  \label{eq20}
\end{eqnarray}

\begin{figure}
{
\begin{center}
	\includegraphics[width=0.7\textwidth]{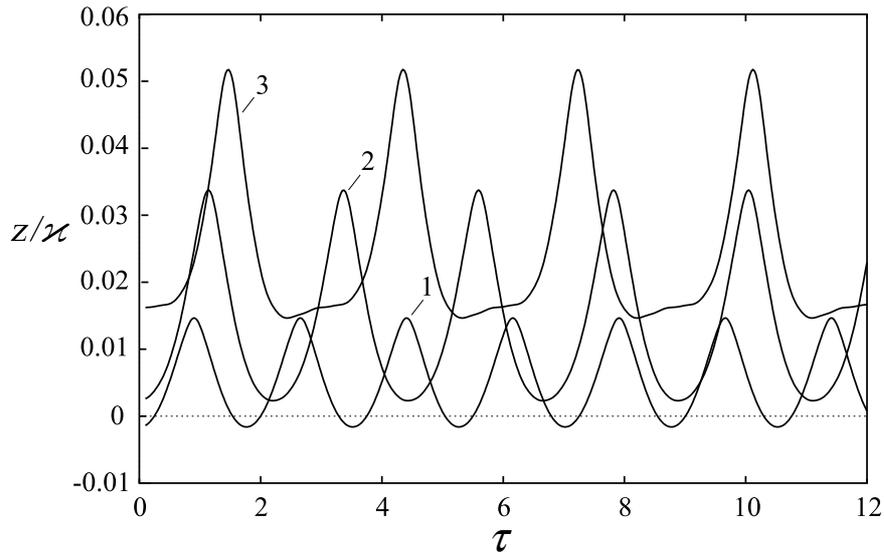}

	\caption{Frequency shift versus time at point $\mathcal{R}=0.1$ for different amplitudes of the halo oscillations: %
	(1) $a_{\max }=0.24$, (2) $ a_{\max }=0.44$, (3) $a_{\max }=0.65$.}
\end{center}
}
\end{figure}

\begin{figure}
{
\begin{center}
	\includegraphics[width=0.7\textwidth]{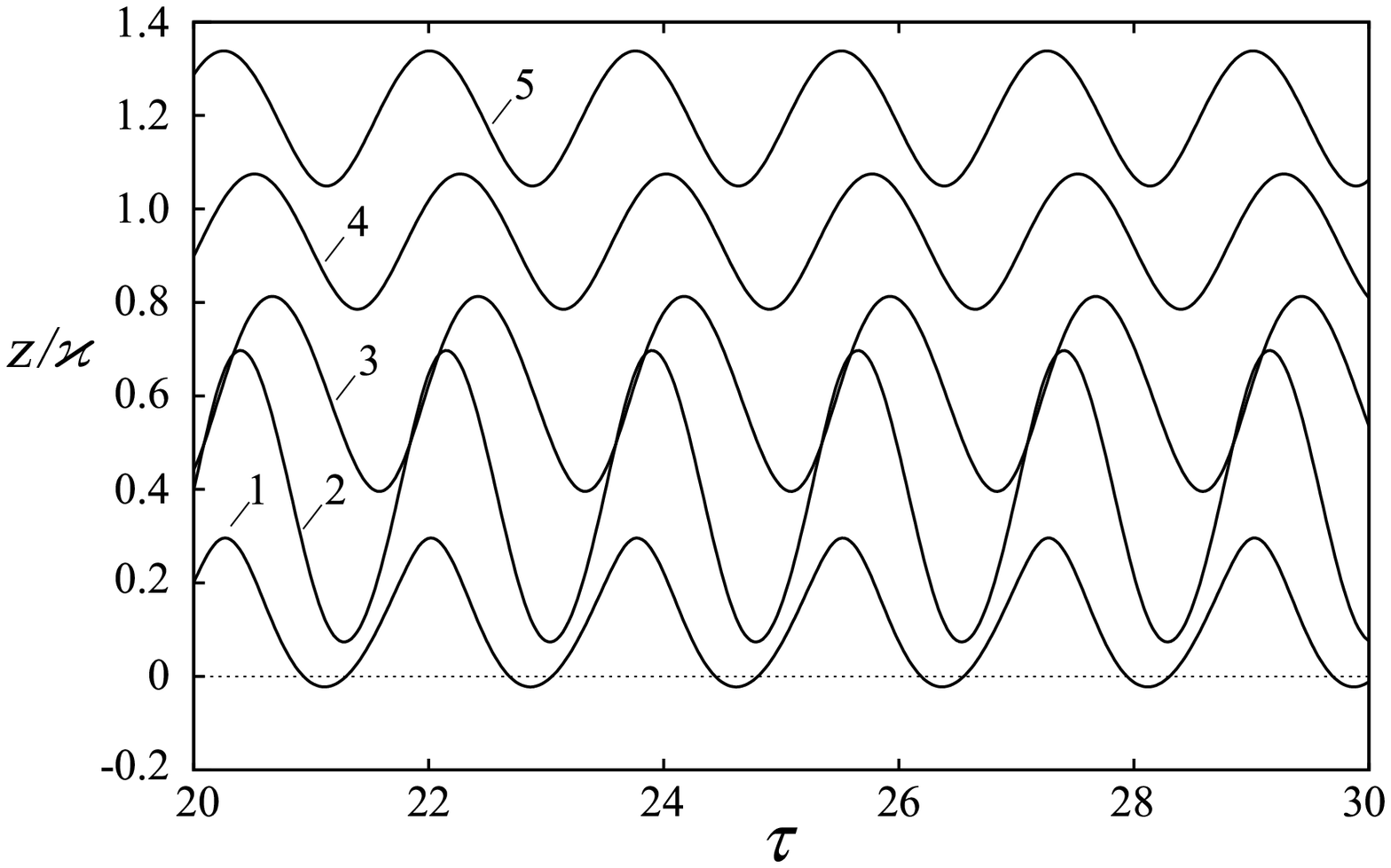}
	
	\caption{Frequency shift versus time for $a_{\max }=0.24$ at different distances from the source: %
	(1) $\mathcal{R}=0.5$, (2) $\mathcal{R}=1$, (3) $\mathcal{R}=1.5$, (4) $\mathcal{R}=3$, (5) $\mathcal{R}=15$.}
\end{center}
}
\end{figure}

At large distances from the source we have%
\begin{eqnarray}
I_{1}(n\omega ,\infty ) &=&n\omega \,e^{-(n\omega )^{2}}{}_{1}F_{1}\left( 
\frac{1}{2};\frac{3}{2};(n\omega )^{2}\right) ,  \nonumber \\
I_{2}(n\omega ,\infty ) &=&\frac{\sqrt{\pi }}{2}e^{-(n\omega )^{2}}, 
\nonumber \\
I_{3}(n\omega ,\infty ) &=&\frac{n\omega }{2}+\frac{n\omega }{2}\big{[}%
1-2(n\omega )^{2}\big{]}\,e^{-(n\omega )^{2}}{}_{1}F_{1}\left( \frac{1}{2};%
\frac{3}{2};(n\omega )^{2}\right) ,  \nonumber \\
I_{4}(n\omega ,\infty ) &=&\frac{\sqrt{\pi }}{4}\big{[}1-2(n\omega )^{2}%
\big{]}\,e^{-(n\omega )^{2}},  \label{eq21}
\end{eqnarray}%
Then in the function $P_{n}$ (\ref{eq18}), in view of (\ref{eq14}), (\ref%
{eq20}), the second square bracket vanishes, so that%
\begin{equation}
P_{n}(\tau (0),\mathcal{\infty })=-\frac{n\omega }{2}A_{n}\cos 2n\omega \tau (0).  
\end{equation}%
As a result, from (\ref{eq17}) we find

\begin{equation}
\left( \frac{\nu _{0}}{\nu _{R}}\right) _{R\rightarrow \infty }\approx 1+\varkappa \frac{e^{3}}{2}\bigg{\{} V_{\max }+\frac{A_{0}}{4}%
+\frac{1}{2}\sum_{n=1}^{\infty }A_{n}\cos 2n\omega \tau (0)\bigg{\}}.  \label{eq22}
\end{equation}

Figures 1 and 2 show the dependence of the frequency shift $z=(\nu _{0}-\nu
_{R})/\nu _{R}$ on the coordinate time at the observation point, calculated
using the formula (\ref{eq17}) for different amplitudes of the halo
oscillations $a_{\max }$ and for different distances $R$ from the source. It
turned out that in view of the rapid convergence of the series, when summing
in (\ref{eq17}), it suffices to take into account only the first 8 Fourier
harmonics.

From the graphs it can be seen that the shift is modulated by the double
frequency of the halo oscillations. The magnitude of the modulations grows
with increasing $a_{\max }$. Note that for some amplitudes and at some
distances, instead of a redshift $(z>0)$, a blue shift $(z<0)$ is observed
at certain time intervals.


\end{document}